\documentclass[twoside,12pt]{article}
\usepackage{epsfig}

\def\Journal#1#2#3#4{{#1} {#2} (#4) #3 }

\def\NPA{{\em Nucl. Phys.} A}

\def\PRL{\em Phys. Rev. Lett.}

\def\PRD{{\em Phys. Rev.} D}

\newcommand{\be}{\begin{equation}}
\newcommand{\ee}{\end{equation}}
\newcommand{\bea}{\begin{eqnarray}}
\newcommand{\eea}{\end{eqnarray}}

\begin{document}
\title{\vspace{1cm} Extracting the Weinberg angle at intermediate energies}
\author{C.\ Praet, N.\ Jachowicz, P.\ Lava, and J.\ Ryckebusch\\
Department of Subatomic and Radiation Physics, Ghent University, Belgium}
\maketitle
\begin{abstract}
A recent experiment by the NuTeV collaboration resulted in a surprisingly high value for the weak mixing angle $\sin^2 \theta_W$.  The Paschos-Wolfenstein relation, relating neutrino cross sections to the Weinberg angle, is of pivotal importance in the NuTeV analysis.  In this work, we investigate the sensitivity of the Paschos-Wolfenstein relation to nuclear structure aspects at neutrino energies in the few GeV range.  Neutrino-nucleus cross sections are calculated for $^{16}$O and $^{56}$Fe target nuclei within a relativistic quasi-elastic nucleon-knockout model. 
\end{abstract}
Neutrinos have astounded physicists on many occasions, in a fascinating process that resulted in the present knowledge about these particles and the weak interaction.  Recently, the NuTeV collaboration extracted an anomalously high value for the Weinberg angle $\sin^2 \theta_W$ \cite{Zeller1}, using  $\nu_{\mu}$ and $\overline{\nu}_{\mu}$ deep inelastic scattering (DIS) reactions on an iron target.  The resulting $\sin^2 \theta_W = 0.22773 \pm 0.00135 \mbox{(stat)} \pm 0.00093 \mbox{(syst)}$ was obtained in an analysis based on the Paschos-Wolfenstein (PW) relation \cite{Paschos1} for an isoscalar target of $u$ and $d$ quarks 
\begin{equation}
\label{eq:PaWo}
\frac{\sigma(\nu,NC) - \sigma(\overline{\nu},NC)}{\sigma(\nu,CC) - \sigma(\overline{\nu},CC)} = \left( \frac{1}{\cos^2 \theta_c} \right)\left(\frac{1}{2} - \sin^2 \theta_W\right).
\end{equation}
Three standard deviations above the common  $\sin^2 \theta_W$ value  of $0.2227 \pm 0.0004$, the NuTeV anomaly predicts a weaker coupling of the neutrino to the $\mbox{Z}^0$-boson.  Whether this result points to new physics or can be explained through a reexamination of the initial analysis, remains a controversial issue \cite{Mohapatra1}.  Accordingly, the NuTeV outcome called for a thorough investigation of the PW relation (\ref{eq:PaWo}).  Although the bulk of proposed explanations addresses QCD uncertainties, considerable attention has been paid to the role played by nuclear effects, e.g. in \cite{Kulagin1}, where the neutron excess correction to the Paschos-Wolfenstein relation is claimed to be larger than the one used in the NuTeV analysis.  

Here, we present a study of the Paschos-Wolfenstein relation for neutrino energies of a few GeV.  Instead of considering (\ref{eq:PaWo}) as a combination of DIS (anti)neutrino-nucleon cross sections,  
quasi-elastic (QE) (anti)neutrino-nucleus cross sections will be employed.  As  a consequence our  PW relation is not longer based on  quark couplings, but is determined by form factor values.  
To calculate the QE cross sections, a fully relativistic nucleon-knockout model \cite{Martinez1} is adopted.  Final-state interactions (FSI) can be included by means of a relativistic Glauber approach, a multiple-scattering extension of the eikonal approximation based on nucleon-nucleon scattering data \cite{Martinez1, Ryckebusch1}.  

\begin{figure}[tb]
\begin{center}
\begin{minipage}[t]{18.5 cm}
\epsfig{file=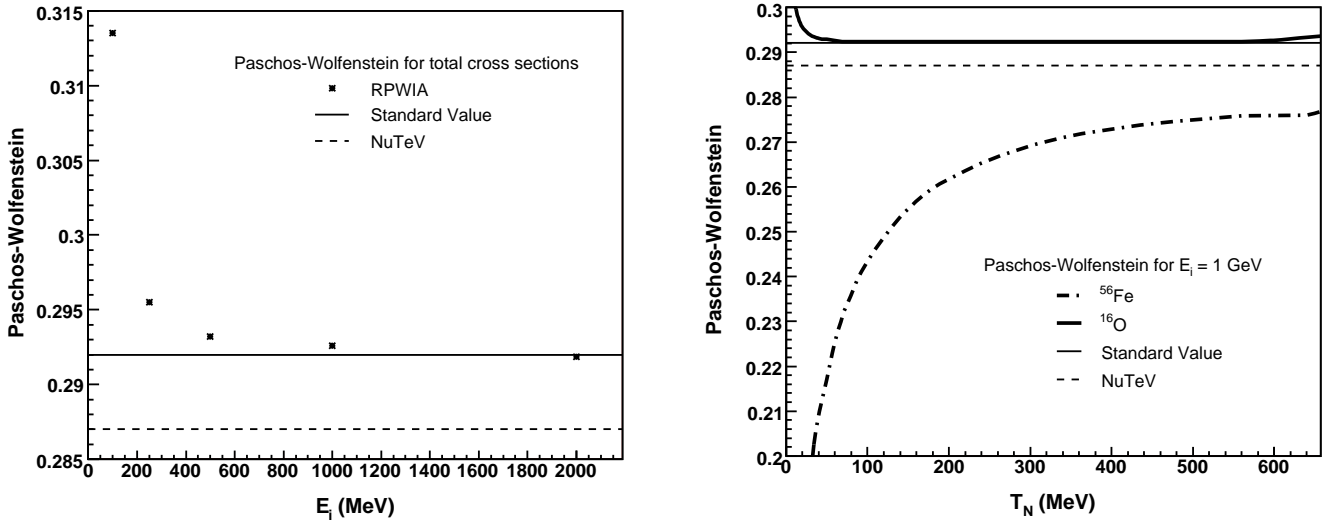,scale=0.925}
\end{minipage}
\begin{minipage}[t]{16.5 cm}
\caption{Comparison between RPWIA results, and the {standard} and NuTeV values for the Paschos-Wolfenstein relation.  Left, total $^{16}$O cross sections for incoming energies ranging from 100 MeV to 2 GeV. The right panel illustrates the dependence of the  Paschos-Wolfenstein relation on the ejectile energy $T_N$ for an incoming neutrino energy of $E_i$ = $1$ GeV for both  $^{16}$O and $^{56}$Fe target nuclei. \label{fig:proc}}
\end{minipage}
\end{center}
\end{figure}

In figure \ref{fig:proc}, we show relativistic plane-wave impulse approximation (RPWIA) results for the Paschos-Wolfenstein relation.  The left panel presents total cross sections
for the isoscalar $^{16}_{\phantom{1}8}$O nucleus.  The effects of FSI are found to be very small. At very low incoming energies, binding effects cause relatively large deviations from the standard PW value (\ref{eq:PaWo}).  For slightly higher incoming energies, the calculated RPWIA results can hardly be distinguished from the standard value, reaching perfect agreement at $E_i = 2$\ \nolinebreak GeV.  Indeed, the $\sin^2 \theta_W$ dependence of the Paschos-Wolfenstein relation, calculated here in a totally different energy regime, is in very good agreement with the standard value ($\sim 0.5 \%$ at $E_i = 500$\ MeV).  

Considering the differential cross sections as a function of the outgoing nucleon's kinetic energy $T_N$, the Paschos-Wolfenstein relation behaves as shown in the right panel of the figure.  The results for $^{16}$O again emphasize the  consistency between our calculation and the standard $\sin^2 \theta_W$ value.  The $^{56}$Fe target on the other hand, yields a different $\sin^2 \theta_W$ dependence.  This can be attributed to the neutron excess in the $^{56}_{26}$Fe nucleus, which strongly enhances the $\sigma(\nu,CC)$ contribution, thereby lowering the PW values quite drastically.  As a consequence, a large neutron-excess correction occurs.  Several other aspects of the Paschos-Wolfenstein relation are currently under study \cite{Praet1}, including the influence of the nucleon's strangeness content.                      

\end{document}